\documentclass[%
 reprint,
 amsmath,amssymb,
 aps,
]{revtex4-2}

\usepackage[normalem]{ulem}
\usepackage{xcolor}
\usepackage{graphicx}
\usepackage{dcolumn}
\usepackage{bm}
\usepackage[utf8]{inputenc}
\usepackage{hyperref}
\hypersetup{
    colorlinks,
    citecolor=blue,
    filecolor=blue,
    linkcolor=blue,
    urlcolor=black
}

\begin{document}

\preprint{APS/123-QED}

\title{Chaotic Chimera Attractors\\ in a Triangular Network of Identical Oscillators}

\author{Seungjae Lee}
 \email{seungjae.lee@tum.de}
\affiliation{Physik-Department, Technische Universit\"at M\"unchen, James-Franck-Stra\ss e 1, 85748 Garching, Germany}

\author{Katharina Krischer}%
 \email{krischer@tum.de}
\affiliation{Physik-Department, Technische Universit\"at M\"unchen, James-Franck-Stra\ss e 1, 85748 Garching, Germany}%

\date{\today}

\begin{abstract}

A prominent type of collective dynamics in networks of coupled oscillators is the coexistence of coherently and incoherently oscillating domains, known as chimera states. Chimera states exhibit various macroscopic dynamics with different motions of the Kuramoto order parameter. Stationary, periodic and quasiperiodic chimeras are known to occur in two-population networks of identical phase oscillators. In a three-population network of identical Kuramoto-Sakaguchi phase oscillators, stationary and periodic symmetric chimeras were previously studied on a reduced manifold in which two populations behaved identically [Phys. Rev. E \textbf{82}, 016216 (2010)]. In this paper, we study the full phase space dynamics of such three-population networks. We demonstrate the existence of macroscopic chaotic chimera attractors that exhibit aperiodic antiphase dynamics of the order parameters. We observe these chaotic chimera states in both finite-sized systems and the thermodynamic limit outside the Ott-Antonsen manifold. The chaotic chimera states coexist with a stable chimera solution on the Ott-Antonsen manifold that displays periodic antiphase oscillation of the two incoherent populations and with a symmetric stationary chimera solution, resulting in tri-stability of chimera states. Of these three coexisting chimera states, only the symmetric stationary chimera solution exists in the symmetry-reduced manifold.

\end{abstract}

\maketitle


\section{\label{sec:intro}Introduction}

Comprehending the dynamics of coupled oscillator ensembles is crucial for various applications, including laser physics~\cite{Larger2015,PhysRevE.96.032215} or Josephson junctions~\cite{josep1,josep2} to biology~\cite{Strogatz1993,winfree2013geometry,10.2307/2101911} and neural science~\cite{Fell2011,Breakspear2017}. Among the fascinating collective behaviors exhibited by identical oscillators is the coexistence of synchrony and asynchrony. This intriguing state was first identified by Battogtokh and Kuramoto~\cite{kuramoto2002} 20 years ago, and was subsequently named a `chimera state'~\cite{abrams2004} to underscore its peculiarity. 

While originally chimera states were investigated in a spatially extended ring geometry~\cite{Panaggio_2015,Omel_chenko_2018,Omel_chenko_2013}, simpler settings, in particular two-population networks revealed essential properties of chimera states~\cite{Kurths_twogroup,abrams_chimera2008}. Here, a chimera state consists of one fully synchronized and one incoherent populations. The success of this model relied on two pillars: First, each oscillator is sinusoidally influenced by an effective force that depends only on the macroscopic dynamics and measures the degree of synchrony of the individual populations~\cite{WS_mobius}. Second, dimension reduction methods allow for the investigation of the dynamics of each population in terms of a few macroscopic variables rather than directly studying the microscopic individual oscillators. In the thermodynamic limit, an invariant manifold, called Ott-Antonsen (OA) manifold~\cite{OA1,OA2}, exists on which the evolution equation for the complex order parameter can be written in a closed form. For finite-sized systems, a so-called Watanabe-Strogatz (WS) transformation~\cite{WS_original1,WS_original2} maps the microscopic dynamics of each population to three macroscopic variables and $N-3$ independent constants of motion~\cite{pikovsky_WS1,pikovsky_WS2,abrams_chimera2016,Bick2020}.

Much about chimera states was learned from two-population networks of identical Kuramoto-Sakaguchi (KS) phase oscillators, arguably the best-studied oscillator model and the simplest network configuration. Three types of stable chimera states were identified: Stationary chimera states where the incoherent population possesses a constant degree of partial coherence, i.e., a stationary magnitude of the Kuramoto order parametere of the incoherent population and breathing chimeras where the degree of partial coherence, respectively the Kuramoto order parameter oscillates periodically~\cite{abrams_chimera2008,abrams_chimera2016,lee1}, and quasiperiodic chimeras, exhibiting quasiperiodic motion of the order parameter~\cite{pikovsky_WS1,pikovsky_WS2}. In contrast, chaotic motion of the macroscopic dynamics seems to require more complex settings, such as the presence of heterogeneities~\cite{sym_twogroup,chaos_kuramoto,hetero_twogroup2}, higher order interactions~\cite{pazo_winfree,Bick_2016_chaotic} or higher-dimensional individual oscillators~\cite{olmi_chaos,Olmi_rotator}. 

Chaotic chimeras, however, are also far less studied, and the settings under which macroscopic chaotic chimeras exist in ensembles of identical KS phase oscillators are not well understood. Note that when we talk here about chaotic chimera states we refer to (macroscopic) chaotic order parameter dynamics in large ensembles of oscillators, not to chaotic (weak) chimeras in systems of just a few, e.g. three~\cite{smallest_chimera} or four oscillators~\cite{sym_twogroup}, where it is not clear to which order parameter dynamics the chaotic motion translates in the thermodynamic limit~\cite{felix}. In this paper, we study what seems to be the simplest topology of a network of identical KS phase oscillators that exhibits, as we will show below, chaotic chimera states: a three-population network of identical KS phase oscillators globally coupled within each population and arranged in a triangle with equal inter-population coupling strengths. 

Three-, or more generally, multi-population networks can be seen as toy models for networks of networks and have been investigated in different contexts. They may exhibit a variety of dynamical states; besides chimera states~\cite{martens_three,martens_three2,laing_ring,three_chaos} and heteroclinic switching between saddle chimeras~\cite{bick_three,Bick2019_m1}, effects of nonresonant natural frequencies~\cite{pikovsky_three}, comparison with three-modal natural frequency distribution~\cite{three_modal}, and the impact of repulsive coupling~\cite{Ott_three,PhysRevE.88.032711} have been studied. 

The macroscopic dynamics of chimera states in three-population networks of identical KS phase oscillators was studied for symmetric solutions where two populations behave identically~\cite{martens_three,martens_three2}. In these studies the thermodynamic limit was considered using the OA approach. Two kinds of chimera states were identified, so called DSD- and SDS-type chimeras, respectively. Here, `S' stands for a completely synchronized population while `D' denotes a desynchronized population. On the symmetry-reduced manifold both of these types of chimeras are stable in some range of the parameters and the degree of coherence of the D-populations can be either stationary or breathing. As the author noted~\cite{martens_three}, nothing can be said about the stability of these symmetric chimeras off the symmetry reduced manifold or about the existence of non-symmetric solutions, in particular of $\text{D}\text{S}\text{D}'$-chimeras, where the $\text{D}'$ indicates that the two desynchronized populations display a different order parameter dynamics. This holds for solutions on the full OA manifold as well as outside it. In this context, macroscopic chaotic chimera states showing aperiodic motion of the order parameter of incoherent populations are of particular interest, as this would be an example of macroscopic chaotic chimeras of identical KS phase oscillators in the thermodynamic limit.

This paper addresses these questions and provides detailed dynamical and spectral properties of observable chimera states. Starting with the microscopic equations, we introduce the order parameter dynamics in both the thermodynamic limit on the full OA manifold and finite-sized ensembles using WS reduction in Sec.~\ref{subsec:governing}. In Sec.~\ref{sec:governing_chaotic_chimera}, we elucidate the coexistence of periodic and chaotic antiphase $\text{D}\text{S}\text{D}'$ chimera attractors on and off the Ott-Antonsen manifold, respectively and demonstrate that these two states also coexist with a symmetric SDS stationary chimera state on the OA manifold. The results are summarized in Sec.~\ref{sec:Conlcusion}.


\section{\label{subsec:governing}Governing Equations}

Consider a system of identical Kuramoto-Sakaguchi phase oscillators, each oscillator $j$ being described by a phase variable $\phi^{(a)}_j(t) \in [-\pi,\pi) =: \mathbb{T}$ where $a=1,2,3$ denotes the population of a three-population network and each population consists of $N$ oscillators.
The 3$N$ microscopic governing equations are given by
\begin{flalign}
    \frac{d}{dt}\phi^{(a)}_j &= \omega + \text{Im}\bigg[ H_a(t)e^{-i\phi_j^{(a)}}\bigg] \notag \\
    &=\omega + \sum_{b=1}^{3}K_{ab} \frac{1}{N}\sum_{k=1}^{N} \sin(\phi_k^{(b)}-\phi_j^{(a)}-\alpha) \label{eq:micro-eq}
\end{flalign} with $j=1,...,N$ and $a,b=1,2,3$. $H_a(t)$ represents a mean-field forcing defined by $H_a(t):= e^{-i\alpha}\big( \mu \Gamma_a(t) + \nu \Gamma_b(t) +\nu \Gamma_c(t) \big)$ where $(a,b,c)$ is a permutation of $(1,2,3)$. $ \Gamma_a(t)$ denotes the complex Kuramoto order parameter of each population, which is defined as 
\begin{equation}
    \Gamma_a(t) = r_a(t)e^{i\Theta_a(t)} := \frac{1}{N}\sum_{j=1}^{N} e^{i \phi^{(a)}_j(t)}. \label{eq:Kuramoto-order-parameter}
\end{equation} The phase-lag parameter $\alpha$ is conveniently written as $\alpha =\frac{\pi}{2}-\beta$. In this work, we fix $\beta=0.025$. Since the oscillators are identical, we can set $\omega = 0$ and the intra-population coupling strength $\mu = 1$. The inter-population coupling strength $\nu$, which we assume to be always weaker than the intra-population coupling strength, is expressed through $\nu=1-A$ where $A \in [0,1]$. Note that the larger the parameter $A$, the weaker the coupling between populations~\cite{martens_three}. In matrix form, the coupling strength of this triangular, symmetric network~\cite{martens_three2} is thus given by
\begin{equation}
   ( K_{a a'}) = \begin{pmatrix}
    \mu &  \nu  & \nu  \\
    \nu & \mu & \nu  \\
    \nu & \nu & \mu  \\ 
    \end{pmatrix} \notag
\end{equation} for $a,a'=1,2,3$. 

Below we analyze the dynamics of Eq.~(\ref{eq:micro-eq}) on the levels of the OA and WS reductions. To derive the global dynamics of the finite-sized system, we exploit the WS transformation~\cite{pikovsky_WS2,WS_original2,WS_mobius}: 
\begin{flalign}
    e^{i\phi_j^{(a)}} = e^{i\Phi_a} \frac{\rho_a + e^{i(\psi_j^{(a)} - \Psi_a)}}{1+\rho_a e^{i(\psi_j^{(a)} - \Psi_a)}} \label{eq:WS-transformation}
\end{flalign} for $j=1,...,N$ and $a=1,2,3$. For each population, the radial variable $\rho_a(t)$ measures the degree of coherence, and $\Phi_a(t)$ the mean phase; note, though, that these quantities are not exactly identical to $r_a(t)$ and $\Theta_a(t)$ of the Kuramoto order parameter. The second angular variable, $\Psi_a(t)$, specifies the motion of the oscillators with respect to the mean phase and $\{ \psi_j^{(a)} \}_{j=1}^{N}$
are $N-3$ independent constants of motion that satisfy three constraints: $\sum_{j=1}^N \cos \psi_j^{(a)} = \sum_{j=1}^N \sin \psi_j^{(a)}=0$ and $\sum_{j=1}^N \psi_j^{(a)} = 0$ for $a=1,2,3$~\cite{WS_original2,pikovsky_WS1,pikovsky_WS2,WS_mobius}. The WS macroscopic variables are related to the complex Kuramoto order parameter via~\cite{pikovsky_WS1,pikovsky_WS2}:
\begin{flalign}
    \Gamma_a(t) = \rho_a(t)e^{i\Phi_a(t)}\gamma_a(\rho_a,\Psi_a;t) \label{eq:relation-kuramoto-WS}
\end{flalign} where $\gamma_a$ is defined by
\begin{flalign}
    \gamma_a &=  \frac{1}{\rho_a} (\zeta_a + i \xi_a) \notag \\ 
    &:= \frac{1}{\rho_a N}\sum_{k=1}^{N}\frac{2\rho_a + (1+\rho_a^2)\cos(\psi_k^{(a)}-\Psi_a)}{1+2\rho_a\cos(\psi_k^{(a)}-\Psi_a)+\rho_a^2} \notag \\
    & +i \frac{1}{\rho_a N}\sum_{k=1}^{N}\frac{ (1-\rho_a^2)\sin(\psi_k^{(a)}-\Psi_a)}{1+2\rho_a\cos(\psi_k^{(a)}-\Psi_a)+\rho_a^2}
\end{flalign} for $a=1,2,3$. With these definitions, the governing equations of the 9D WS variables are given as~\cite{pikovsky_WS1,pikovsky_WS2,abrams_chimera2016}
\begin{flalign}
    \frac{d\rho_a}{dt} &= \frac{1-\rho_a^2}{2}\sum_{a'=1}^{3}K_{a a'}\big( \zeta_{a'} \cos(\Phi_{a'}-\Phi_a-\alpha) \notag \\
    &- \xi_{a'} \sin(\Phi_{a'}-\Phi_a-\alpha) \big), \notag \\
    \frac{d\Psi_a}{dt} &= \frac{1-\rho_a^2}{2\rho_a}\sum_{a'=1}^{3}K_{a a'}\big( \zeta_{a'} \sin(\Phi_{a'}-\Phi_a-\alpha) \notag \\
    &+ \xi_{a'} \cos(\Phi_{a'}-\Phi_a-\alpha) \big),  \notag \\ \frac{d\Phi_a}{dt} &= \frac{1+\rho_a^2}{2\rho_a}\sum_{a'=1}^{3}K_{a a'}\big( \zeta_{a'} \sin(\Phi_{a'}-\Phi_a-\alpha) \notag \\
    &+ \xi_{a'} \cos(\Phi_{a'}-\Phi_a-\alpha) \big) \label{eq:WS-governing-eq}
\end{flalign} for $a=1,2,3$.

The OA dynamics can be obtained from the WS dynamics for uniform constants of motion $\psi_j^{(a)}=-\pi + \frac{2\pi (j-1)}{N}$ for $j=1,...,N$, taking the thermodynamic limit $N \rightarrow \infty$. Under this condition, the Kuramoto order parameter is exactly described by $\Gamma_a(t) = \rho_a(t)e^{i\Phi_a(t)}$ since $\gamma_a = 1$ for $a=1,2,3$ and the governing equations read~\cite{martens_three,pikovsky_WS1,pikovsky_WS2}
\begin{flalign}
    \frac{d\rho_a}{dt} &= \frac{1-\rho_a^2}{2}\sum_{a'=1}^{3}K_{aa'}\rho_{a'}\cos(\Phi_{a'}-\Phi_a-\alpha), \notag \\
    \frac{d\Phi_a}{dt} &= \frac{1+\rho_a^2}{2\rho_a}\sum_{a'=1}^{3}K_{aa'}\rho_{a'}\sin(\Phi_{a'}-\Phi_a-\alpha) \label{eq:6D-OA-governing}
\end{flalign} for $a=1,2,3$.

\section{\label{sec:governing_chaotic_chimera}Symmetry-broken Chimeras}

\begin{figure}[t!]
\includegraphics[width=1.0\linewidth]{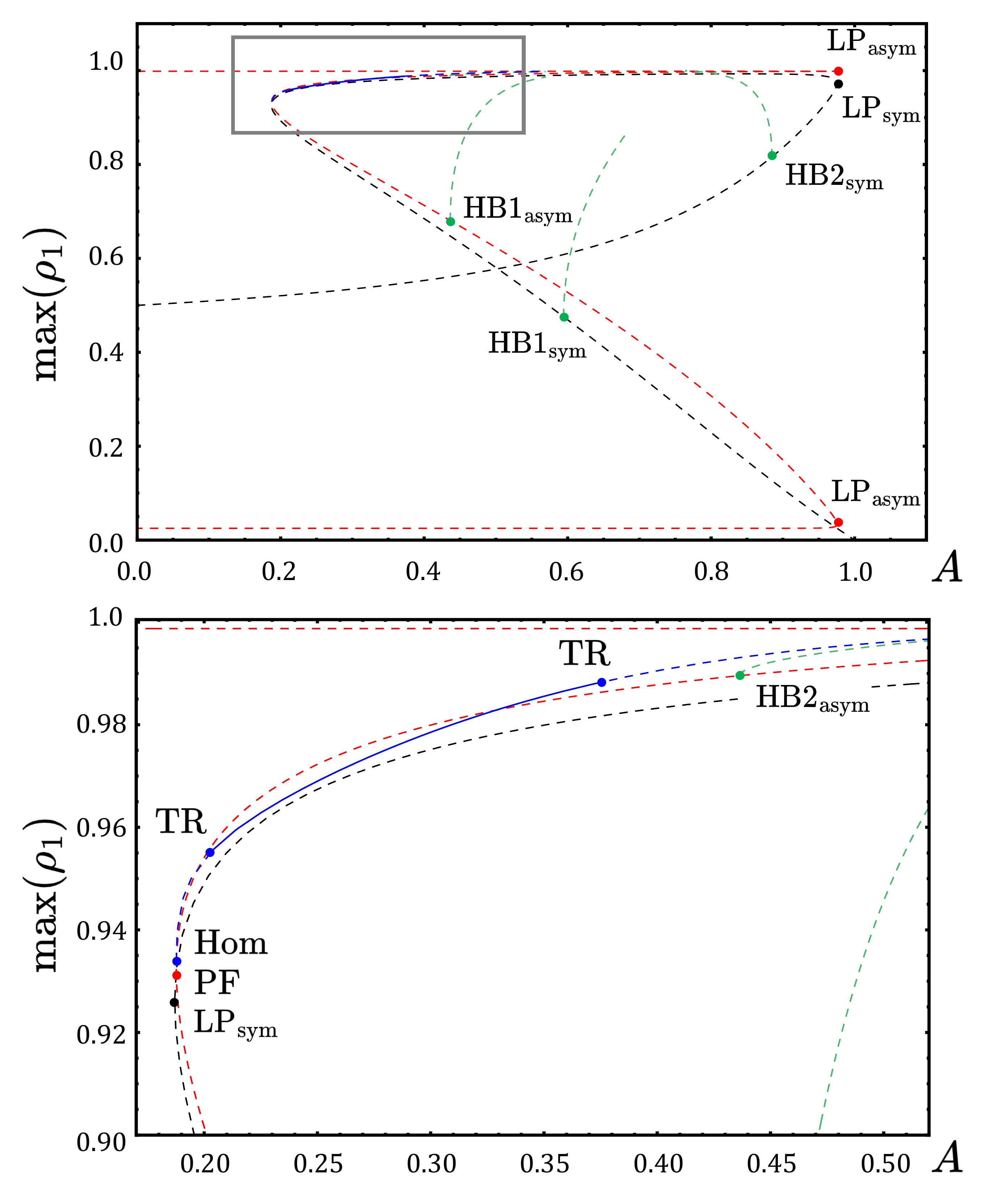}
\caption{Upper plate: Bifurcation diagram of DSD-type chimera states. Lower plate: Enlargement of the gray box in the upper plate. Dashed and solid lines indicate unstable and stable curves, respectively. Black: symmetric stationary DSD. Red: asymmetric stationary $\text{D}\text{S}\text{D}'$. Green: symmetric breathing chimeras. Blue: antiphase $\text{D}\text{S}\text{D}'$ chimera states.} 
\label{Fig:DSD-bif}
\end{figure}

We first focus our attention on observable, i.e., stable chimera states in the thermodynamic limit which live on the Ott-Antonsen manifold. As mentioned above, there are two basic types of chimera states, SDS- and DSD-types chimeras. The detailed analysis of the symmetric SDS and DSD chimera states, which do not lead to a chaotic chimera state, are given in Appendix.~\ref{sec:other_chimeras}. Here, we concentrate on the symmetry-broken $\text{D}\text{S}\text{D}'$ chimera states, which are connected to the chaotic chimeras. 

\begin{figure}[t!]
\includegraphics[width=1.0\linewidth]{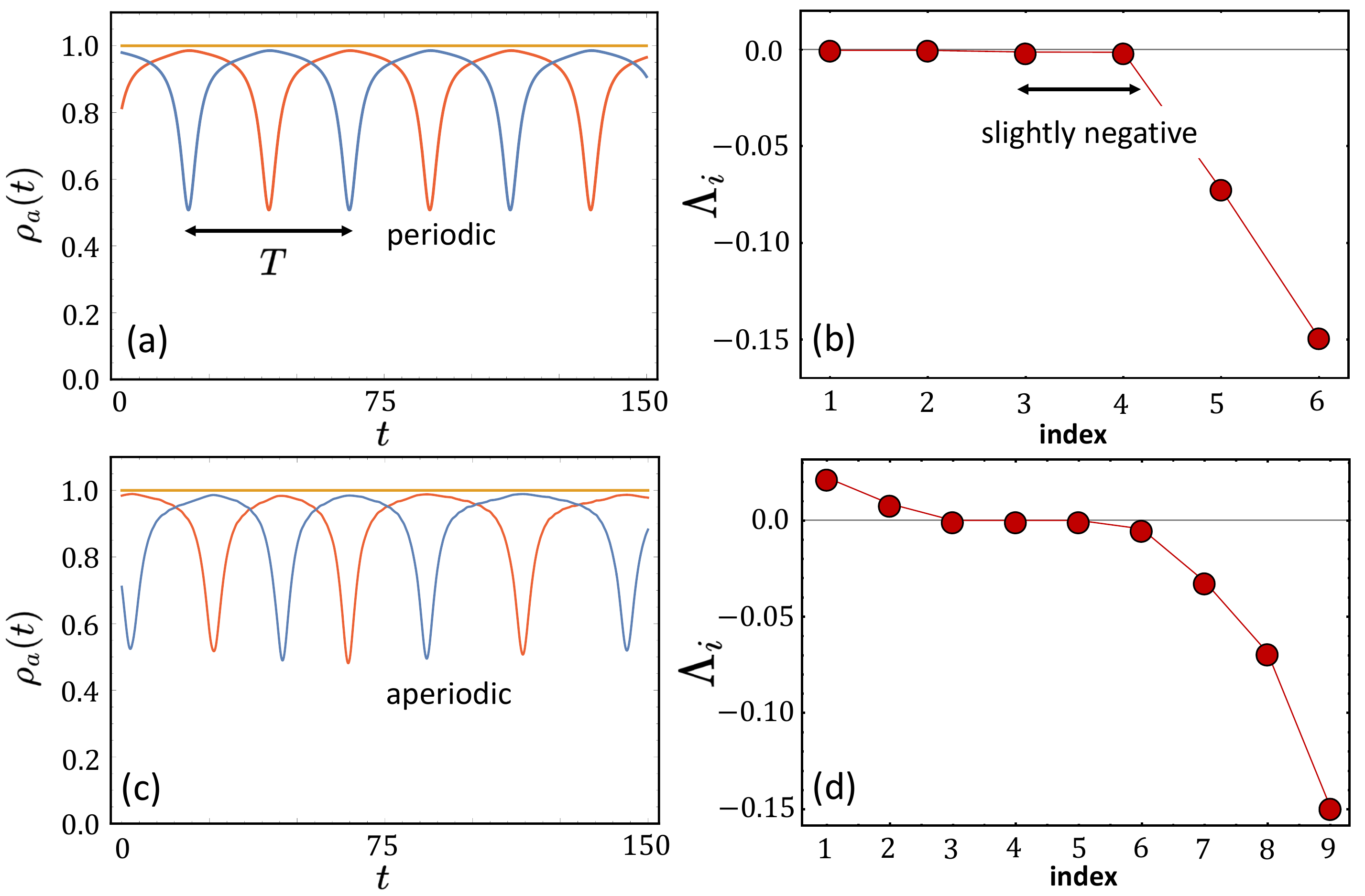}
\caption{(a) Radial variables of the 6D OA dynamics (red: first, blue: third population, and orange: second population) and (b) Lyapunov exponents along the antiphase chimera trajectory. (c) Radial variables of the 9D WS dynamics with uniform constants of motion and $N=20$ and (d) and the corresponding Lyapunov exponents. In subfigures: $A=0.35$ and all are measured after $t> 10^5$.} 
\label{Fig:antiphase_OA}
\end{figure}

A bifurcation diagram of DSD-type chimera states is depicted in Fig.~\ref{Fig:DSD-bif}. The black curves correspond to the symmetric DSD chimeras with $\rho_1(t) = \rho_3(t)<1$ while $\rho_2(t)=1$ and $\Phi_1(t) = \Phi_3(t) \in \mathbb{T}$ that live in the symmetry-reduced manifold. Their dynamics within this manifold was studied in Ref.~\cite{martens_three}. Our results are in agreement with these data, but reveal that all symmetric DSD chimera states, i.e. also those that were found to be stable within the reduced manifold, have at least one transversally unstable direction, which drives the two symmetric $\text{D}$-populations in opposite directions, and are thus unstable (see Appendix.~\ref{subsec:DSD} for details). Not reported before are the asymmetric $\text{D}\text{S}\text{D}'$ chimeras with $\rho_1(t) \neq \rho_3(t) <1, \Phi_1(t) \neq \Phi_3(t)$ that live off the reduced manifold. Among these states are stable asymmetric $\text{D}\text{S}\text{D}'$ chimera states (blue) in which the two incoherent populations exhibit antiphase character: $\rho_1(t) = \rho_3(t-\frac{T}{2})$ where $T$ is the period of the oscillation, and thus, $\rho_1(t)\neq \rho_3(t)$ for $t \in \mathbb{R}_{+}$ almost everywhere. Time series of the three moduli of the order parameters are shown in Fig.~\ref{Fig:antiphase_OA} (a). These antiphase $\text{D}\text{S}\text{D}'$ chimeras are born in a double-homoclinic cycle of the stationary symmetric DSD chimeras, which is point-symmetric in the projection on the $\rho_1 \rho_3$-plane with respect to the diagonal line ($\rho_1=\rho_3$). The homoclinic bifurcation (Hom) lies very close to a pitchfork bifurcation (PF) at which a pair of stationary $\text{D}\text{S}\text{D}'$ chimeras (red) bifurcates off the stationary DSD chimera state (black). The homoclinic bifurcation throws off unstable antiphase $\text{D}\text{S}\text{D}'$ chimeras (dashed blue) that are at somewhat larger values of $A$ stabilized in a subcritical torus bifurcation (TR). Further increasing $A$, the antiphase chimera state eventually becomes unstable again via a subcritical torus bifurcation. 

Though the antiphase $\text{D}\text{S}\text{D}'$ chimera state results from the broken symmetry due to the transversal instability, the dynamics of them inherits another symmetry of the solution: $\dot{\Phi}_a(t)=\dot{\Phi}_a(t-T),\dot{\rho}_a(t)=\dot{\rho}_a(t-T)$ for $a=1,3$ and $\dot{\Phi}_2(t)=\dot{\Phi}_2(t-\frac{T}{2}),\dot{\Phi}_1(t)=\dot{\Phi}_3(t-\frac{T}{2})$ which leads to $\rho_1(t) = \rho_3(t-\frac{T}{2})$. Thus, antiphase chimera states possess just 2 effective degrees of freedom. This fact is also reflected in the spectrum of the Lyapunov exponents (LEs) (Fig.~(\ref{Fig:antiphase_OA}) (b))~\cite{pikovsky_LE,CLV1,CLV2}. There are two zero Lyapunov exponents due to continuous symmetries, i.e., time and phase shift invariance, with two slightly negative and two prominently negative Lyapunov exponents. Hence, the antiphase $\text{D}\text{S}\text{D}'$ chimeras in the OA manifold are a stable periodic motion, i.e., a periodic antiphase chimera.

\begin{figure}[t!]
\includegraphics[width=1.0\linewidth]{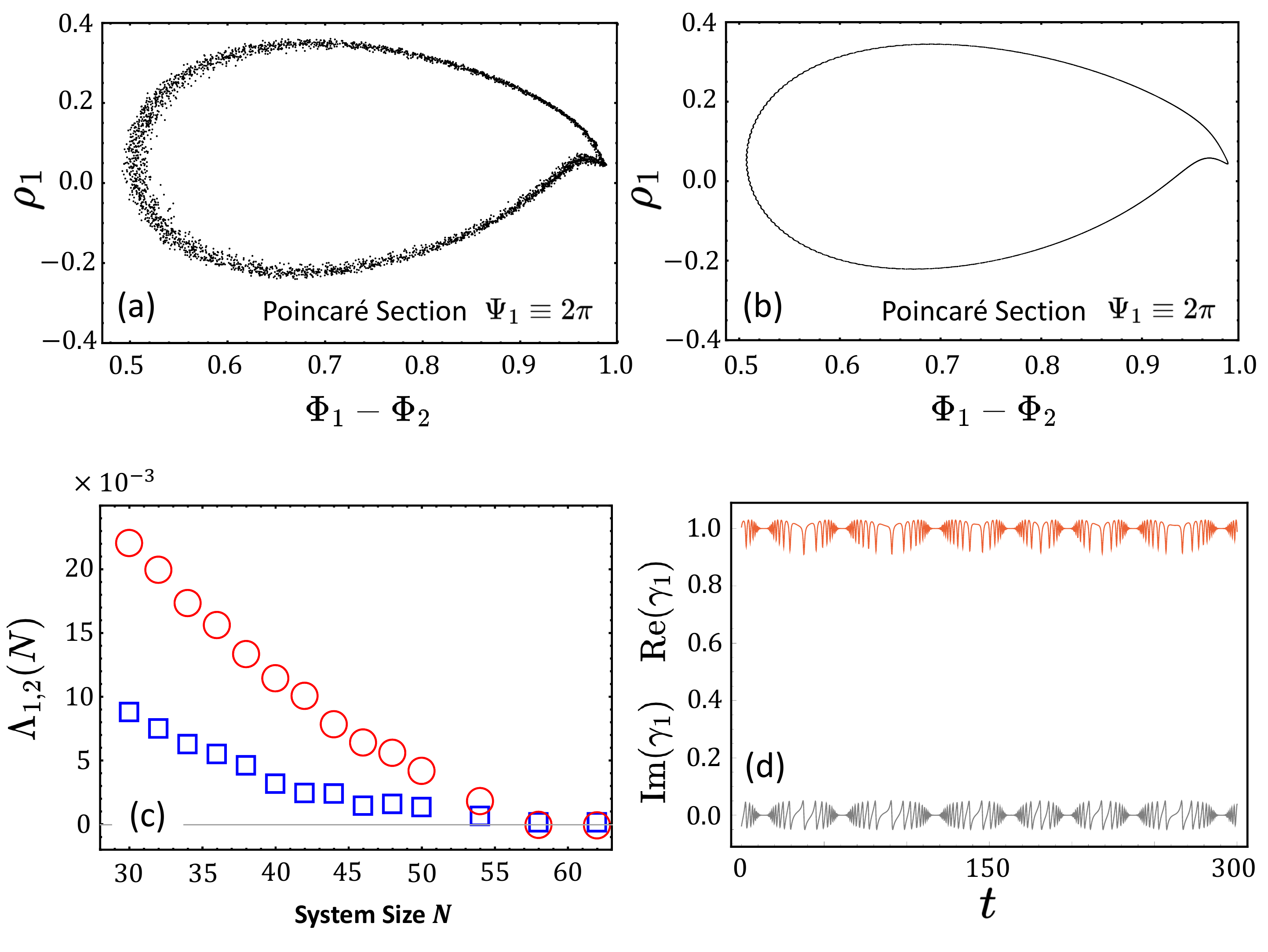}
\caption{ Poincar\'e section of 9D WS dynamics with uniform constants of motion and $\Psi_1 \equiv 2\pi$ for (a) $N=30$ and (b) $N=100$. (c) The largest and the second largest LEs as a function of $N$ for 9D WS dynamics. (d) Real (orange) and imaginary (gray) parts of $\gamma_1$ as a function of time after $t>10^4$. All the results are obtained with uniform constants of motion of WS dynamics, and $A=0.35$.} 
\label{Fig:antiphase_WS}
\end{figure}

In Fig.~\ref{Fig:antiphase_OA} (c), the antiphase chimera state in a finite-sized system obtained from the WS dynamics (\ref{eq:WS-governing-eq}) with uniform constants of motion is shown. Although it displays oscillating antiphase motion of the two incoherent populations similar to the one of the OA dynamics, its motion is not periodic but rather aperiodic, and thus the state does not possess the above mentioned spatio-temporal symmetry, i.e., $\rho_1(t) \neq \rho_3(t-\frac{T}{2})$. In fact, the Lyapunov analysis (Fig.~\ref{Fig:antiphase_OA} (d)) clearly shows two positive Lyapunov exponents. In other words, the antiphase chimera state of the small-sized system in the Poisson submanifold (defined as a manifold of finite size $N$ as close as possible to the OA manifold, and obtained with uniform constants of motion) is chaotic. Furthermore, besides the evidence of the two positive LEs, the chaotic motion of the WS dynamics in small-$N$ systems can be detected in a Poincar\'e section defined by $\Psi_1 \equiv 2\pi$ (Fig.~\ref{Fig:antiphase_WS} (a))~\cite{abrams_chimera2016}. For $N=30$, the dynamics in the section follows a scattered motion on a band-like region, as expected for motion on a chaotic attractor.

\begin{figure}[t!]
\includegraphics[width=1.0\linewidth]{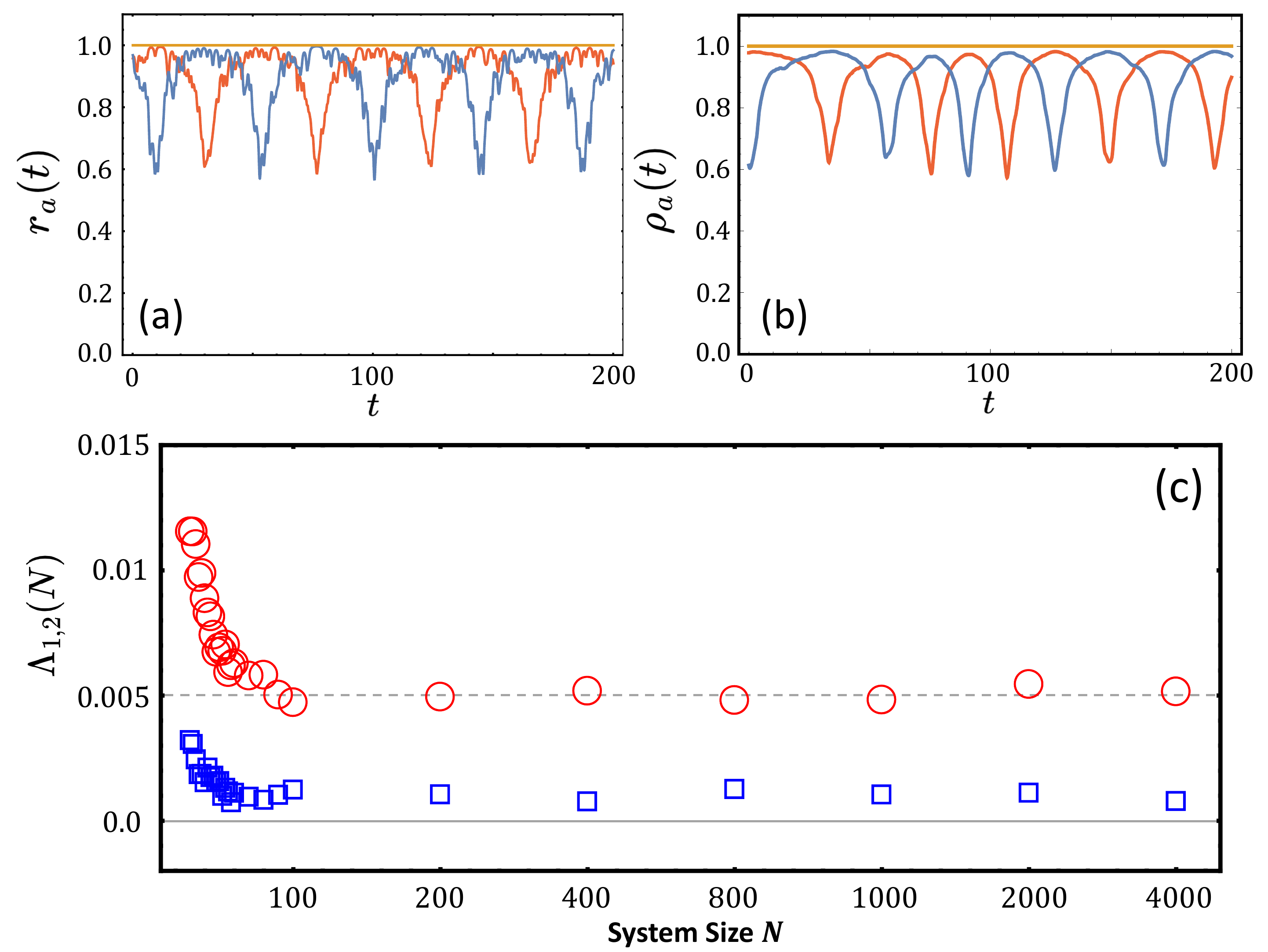}

\caption{(a) Moduli of the Kuramoto order parameters of the three populations for the microscopic dynamics (\ref{eq:micro-eq}) with $N=40$ after $t>10^5$. (b) Radial variables of the WS dynamics (\ref{eq:WS-governing-eq}) with $N=40$ after $t>10^5$. (c) The largest and the second largest LEs as a function of $N$. The 9D WS dynamics is obtained with nonuniform constants of motion and the microscopic dynamics from a random initial condition. The parameter used here is $A=0.3$.  } 
\label{Fig:antiphase_WS_nonuniform}
\end{figure}

Opposed to this, the dynamics of the large system with $N=100$ oscillators resides on a one-dimensional curve in the Poincar\'e section (Fig.~\ref{Fig:antiphase_WS} (b)) suggesting that the chaotic motion is restricted to small system sizes. This conjecture is further supported by calculations of the two largest Lyapunov exponents as a function of system size $N$. 
In Fig.~\ref{Fig:antiphase_WS} (c) we can see that the two positive Lyapunov exponents decrease until $N \approx 60$ where they become numerically indistinguishable from zero. The deterministic small-size effect that gives rise to a chaotic motion of antiphase chimeras arises from the influence of $\gamma_a \in \mathbb{C}$ on the WS variables. For small $N$, $\gamma_a$ significantly affects the dynamics given by Eq.~(\ref{eq:WS-governing-eq}) and renders the WS dynamics different from the OA dynamics. The irregular motion of the real and the imaginary part of $\gamma_a$ as a function of time of the chaotic chimera state depicted in Fig.~\ref{Fig:antiphase_WS} (a) is displayed in Fig.~\ref{Fig:antiphase_WS} (d). As $N$ approaches infinity, $\text{Re}(\gamma_a) \rightarrow 1$ and $\text{Im}(\gamma_a) \rightarrow 0$, so that Eq.~(\ref{eq:WS-governing-eq}) becomes identical to Eq.~(\ref{eq:6D-OA-governing}) and the aperiodic WS dynamics coincides with the periodic OA motion.

Off the OA manifold, the picture is different. First, we observe that starting from random initial conditions picked from $\mathbb{T}^{3N}$ for simulations of the microscopic dynamics (\ref{eq:micro-eq}), chaotic antiphase chimera states are observed as well. An example is shown in Fig.~\ref{Fig:antiphase_WS_nonuniform} with $N=40$ where the evolution of the moduli of the Kuramoto order parameters for the microscopic dynamics (Fig.~\ref{Fig:antiphase_WS_nonuniform} (a)) is depicted together with the radial variables of the WS dynamics (Fig.~\ref{Fig:antiphase_WS_nonuniform} (b)). The latter is obtained from nonuniform constants of motion that are generated using $\psi_j^{(a)} = (1-q)\frac{\pi}{2} + \frac{\pi q (j-1)}{N/2}$ and $\psi_{j+N/2}^{(a)} = -(1+q)\frac{\pi}{2} + \frac{\pi q (j-1)}{N/2}$ with $q=0.85$ for $j=1,...,\frac{N}{2}$ and for $a=1,2,3$~\cite{pikovsky_WS1,pikovsky_WS2}. The moduli of the Kuramoto order parameters exhibit a qualitatively similar envelope to the radial variables, but with further fluctuations superimposed, as expected from the impact of $\gamma_a$ in Eq.~(\ref{eq:relation-kuramoto-WS}). However, different from the dynamics in the OA manifold, outside of it, the chaotic motion persists for systems as large as $N=4,000$. The two largest Lyapunov exponents of simulations with nonuniformly distributed constants of motion are shown in Fig.~\ref{Fig:antiphase_WS_nonuniform} (c). Up to about $N=100$ they decrease with system size, but then saturate at a positive, non-zero value, indicating that outside the OA manifold the chaotic attractors exist even in the thermodynamic limit. Likewise, the Poincar\'e section elucidates a similar scattered characteristics to Fig.~\ref{Fig:antiphase_WS} (a) even for $N=200$ with nonuniform constants of motion (not shown here). Note that this chaotic motion is not microscopic chaos but rather it is a macroscopic chaos of the order parameter dynamics~\cite{nakagawa1,nakagawa2}. Consequently, we show that in the thermodynamic limit, a system of identical Kuramoto-Sakaguchi phase oscillators in three-population networks supports the coexistence of periodic antiphase chimeras within the OA manifold and chaotic antiphase chimeras outside of it. These symmetry-broken chimeras arise due to the transversal instability of symmetric DSD chimeras between two symmetric populations. Above we discussed that in the thermodynamic
limit, three populations support the coexistence
of periodic antiphase chimeras within the OA manifold
and chaotic antiphase chimeras outside it. In fact, in
addition to these two, the stationary SDS chimera state
coexists with these two states in a wide parameter range (see Appendix.~\ref{subsec:SDS} for details of stability of the symmetric SDS chimeras). Therefore, we evidence the coexistence of three observable chimera states. In addition, the fully synchronized solution is stable in the entire parameter range in which chimera states exist. Yet, from 200 simulations from random initial condition within and off the OA manifold each, at e.g. $A=0.3$ about half of them lead to antiphase chimeras (periodic in the OA and chaotic off the OA) and the other half of them to symmetric SDS chimeras (see Appendix.~\ref{subsec:SDS}), evidencing that the basins of the three coexisting chimera states occupy most of the phase space.

\section{\label{sec:Conlcusion}Summary}

Our studies revealed that in three-population networks of identical Kuramoto-Sakaguchi phase oscillators, a variety of qualitatively different chimera states can not only exist, but also coexists and jointly attract most of the initial conditions in phase space. 

Previously, symmetric chimera states, namely SDS and DSD-chimeras with two populations behaving alike, were reported to exist in the symmetry-reduced manifold~\cite{martens_three}. In this work, we have lifted the symmetry constraint and studied the full dynamics at the level of both microscopic dynamics and macroscopic dynamics, using the Watanabe-Strogatz and the Ott-Antonsen approaches. In full phase space, the symmetric DSD chimeras are unstable to transversal perturbations, i.e. perturbations in which the two D-populations move in opposite directions, resulting in $\text{D}\text{S}\text{D}'$ states. These asymmetric $\text{D}\text{S}\text{D}'$ chimeras are stable in a wide parameter range. They form chaotic antiphase chimera attractors in both finite-sized systems and in the thermodynamic limit outside the Ott-Antonsen manifold. In the OA manifold, such an antiphase chaotic chimera is rendered periodic. Hence, these two types of antiphase $\text{D}\text{S}\text{D}'$ chimeras coexist in the thermodynamic limit. Furthermore, these two types of chimeras coexist with a symmetric stationary SDS chimera state in apparently the entire parameter range of their existence. The simple transition from two- to three-population networks results thus in a quantum jump in the richness of dynamics, and in particular supporting tri-stability of chimera attractors, whereby the three coexisting states encompass stationary, oscillating and chaotic chimeras.

This work illustrates in detail how rich the order parameter dynamics of oscillators with identical natural frequencies in more complex network topologies can be. It thus complements and supports recent results on M-population networks of coupled heterogeneous oscillators~\cite{laing_ring} and will certainly lead to further investigations in the future.

\begin{acknowledgments}
The authors would like to thank Young Sul Cho for providing additional computing facility. This work has been supported by the Deutsche Forschungsgemeinschaft (project KR1189/18-2).
\end{acknowledgments}

\appendix

\section{\label{sec:other_chimeras}Symmetric Chimeras}

\begin{figure}[t!]
\includegraphics[width=1.0\linewidth]{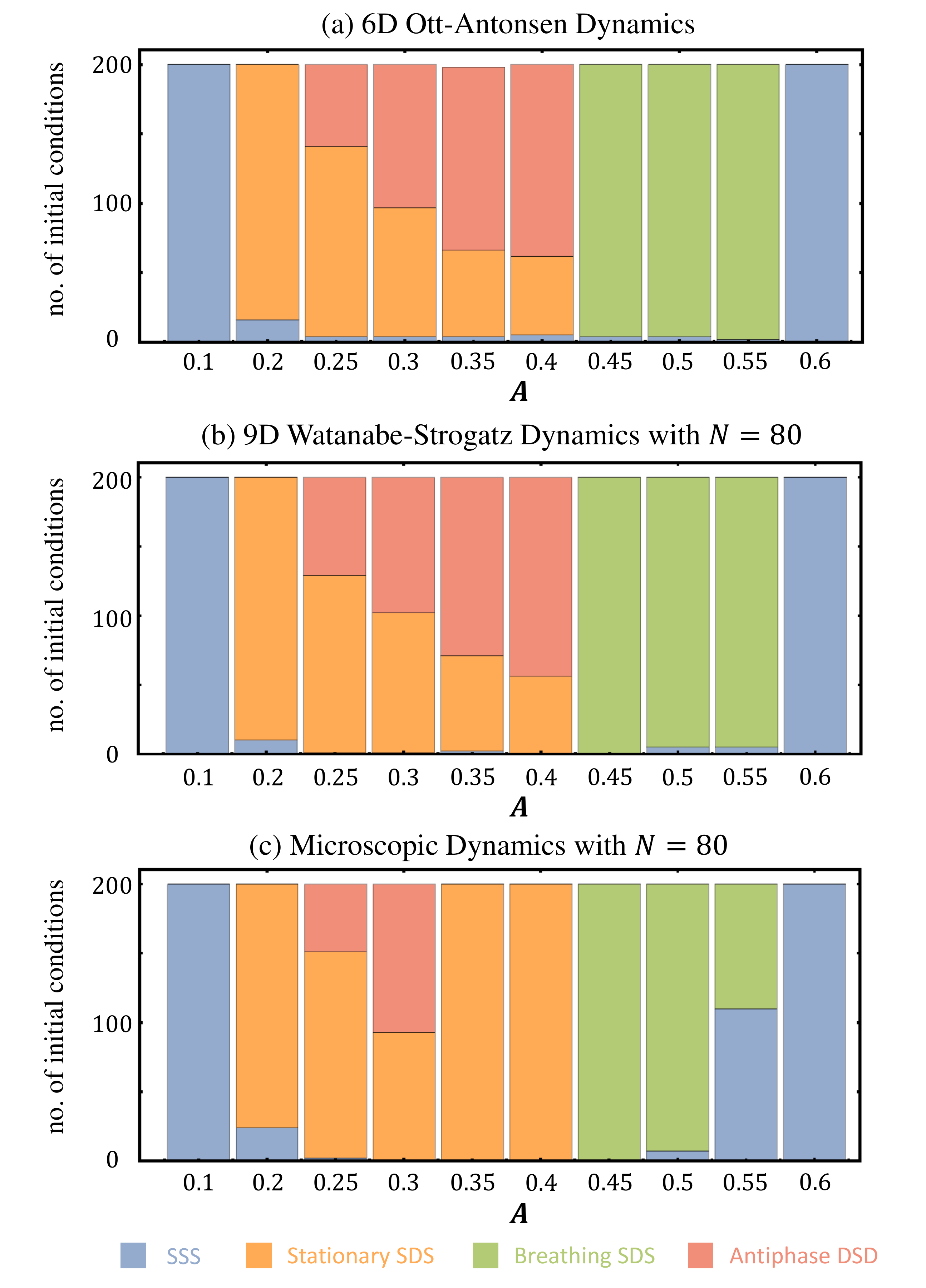}
\caption{The number of dynamical states at $t=20,000$ starting from 200 random initial conditions for different values of $A$. (a) 6D Ott-Antonsen dynamics. (b) 9D Watanabe-Strogatz macroscopic dynamics with $N=80$ and uniform constants of motion. (c) $3N$-dimensional microscopic dynamics with $N=80$.}
\label{Fig:ensemble-table}
\end{figure}

In order to determine the likeliness of observing periodic and chaotic antiphase chimera states, we performed 200 simulations with the OA dynamics (Eq.~(\ref{eq:6D-OA-governing})), the WS dynamics (Eq.~(\ref{eq:WS-governing-eq})) and the microscopic dynamics (Eq.~(\ref{eq:micro-eq})), respectively. For the WS dynamics, we used uniform constants of motion for $N=80$. They are all initialized with random initial conditions of the corresponding dynamical variables. The outcome is shown in Fig.~\ref{Fig:ensemble-table}. In the range between $0.2 \lesssim A \lesssim 0.4$, the coexistence of the symmetric stationary SDS chimeras and the antiphase $\text{D}\text{S}\text{D}'$ chimeras is observed in the OA and WS dynamics. Here, the antiphase chimeras are periodic. From the microscopic dynamics which corresponds to nonuniform constants of motion in the WS dynamics, chaotic antiphase chimeras are coexisting with symmetric stationary SDS chimera states, though in a somewhat smaller parameter interval. Besides these, also symmetric breathing SDS chimeras are found, though at different values of $A$. Note that in Ref.~\cite{martens_three}, the authors considered only solutions on the symmetry-reduced manifold where the populations one and three behave alike in the Ott-Antonsen dynamics. They observed stable symmetric stationary/breathing SDS and DSD chimera states in the reduced manifold. However, the symmetric stationary and breathing DSD chimera states are unstable in the entire phase space.

\subsection{\label{subsec:SDS}Symmetric Chimera States of SDS-type}

SDS-type symmetric chimera states in three-population networks behave similarly to those in two-population networks~\cite{lee1,abrams_chimera2008,abrams_chimera2016}. In this section, we discuss their dynamical and spectral properties.
\begin{figure}[t!]
\includegraphics[width=1.0\linewidth]{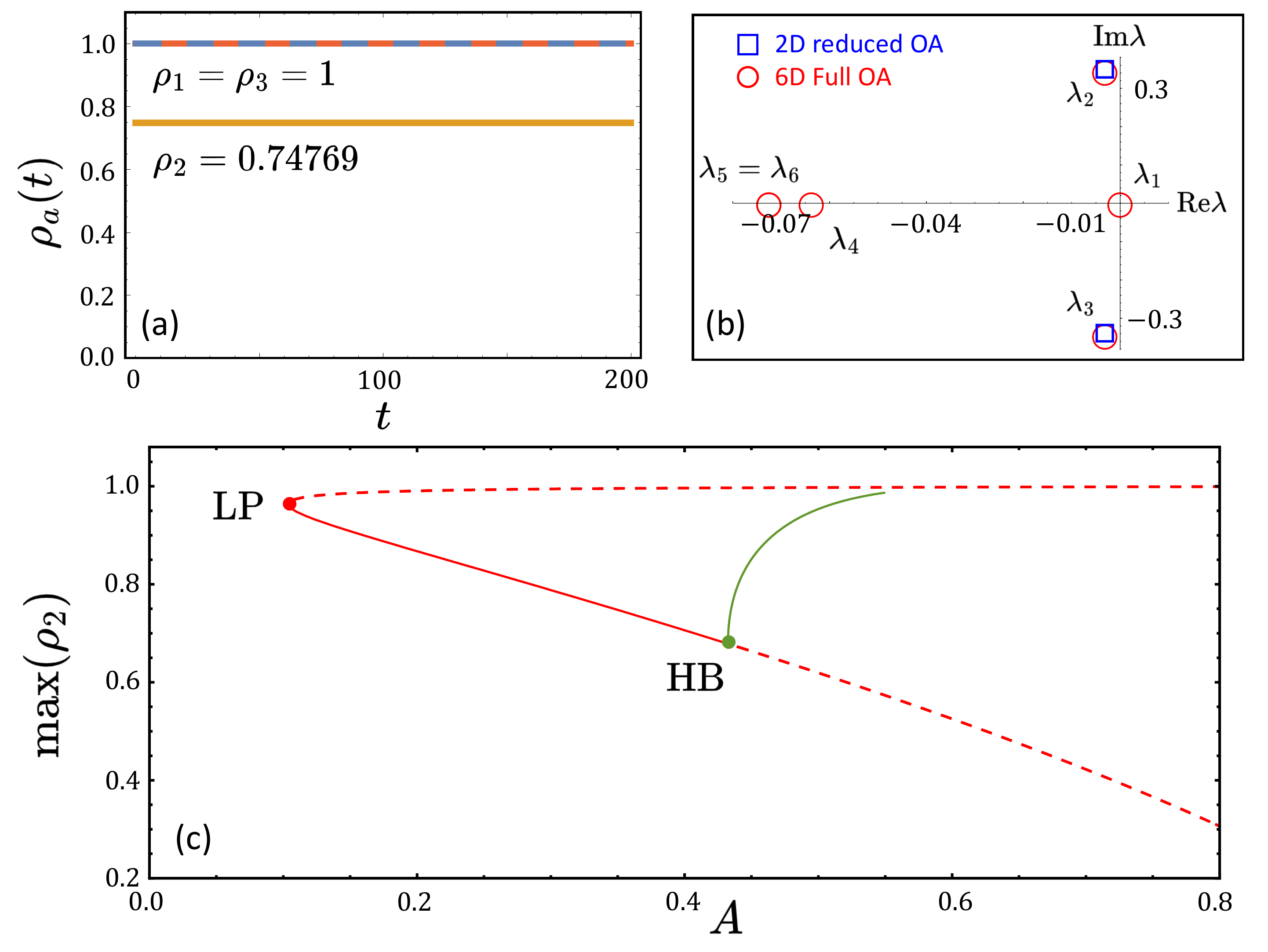}
\caption{ Stationary SDS chimeras in the 6D OA dynamics. (a) Time evolution of the radial variables of 6D OA dynamics after a transient time of $10^5$ units for $A=0.35$. (b) The eigenvalues of the Jacobian matrix evaluated at the stationary SDS chimera fixed point solution shown in (a) in the complex plane. Red circles indicate the eigenvalues in the 6D OA dynamics, and the blue squares those in the 2D symmetry-reduced manifold. (c) Bifurcation diagram of SDS chimeras. The states are born in a limit point bifurcation (LP). The red dashed and solid curves indicate the location of unstable and stable stationary SDS chimeras, respectively. The green curve shows the maxima of the radial variable of a breathing chimera state emerging in a superciritical Hopf bifurcation (HB). } 
\label{Fig:SDS-OA}
\end{figure}
First,  we investigate the stationary SDS chimera states in the 6D Ott-Antonsen manifold. This solution is characterized by $\rho_1(t)=\rho_3(t)=1$, $\rho_2(t) = \rho_0 <1$, $\partial_t \Phi_a(t) = \Omega \in \mathbb{R}$ for $a=1,2,3$. Figure~\ref{Fig:SDS-OA} (a) shows  $\rho_a(t)$ of a stationary SDS at $A=0.35$. The angular velocity is numerically found to be $\Omega =-2.13448$. In a frame rotating with $\Omega$, this solution can be viewed as a fixed point. Hence we can perform a linear stability analysis and determine the eigenvalues of the Jacobian matrix evaluated at the fixed point solution. In Fig.~\ref{Fig:SDS-OA} (b), the six eigenvalues are plotted in the complex plane. All the eigenvalues have non-positive real parts. The eigenvalue, $\lambda_1=0$ has an accompanying eigenvector given by $\delta x_1 = (0,0,0,\delta a,\delta a,\delta a)^\top$ where $\delta a = 1/\sqrt{3}$, which results from the phase shift invariance and does not affect the stability. Thus, the fixed point solution is stable. The pair of complex conjugate eigenvalues, $\lambda_2 = \overline{\lambda}_3$ corresponds to the eigenvector within the symmetry-reduced manifold: $\delta x_2 = \overline{\delta x}_3 = (0,\delta a,0, \delta b, \delta c, \delta b)^\top$ for $\delta a, \delta c \in \mathbb{C}$ and $\delta b \in \mathbb{R}$. The real negative eigenvalue $\lambda_4$ is related to the perturbation transverse to the symmetry-reduced manifold in the angular direction: $\delta x_4 = (0,0,0,\delta a, 0, -\delta a)^\top$ where $\delta a  = 1/\sqrt{2}$. The last two eigenvalues $\lambda_5=\lambda_6$ are real negative and degenerate.

\begin{figure}[t!]
\includegraphics[width=1.0\linewidth]{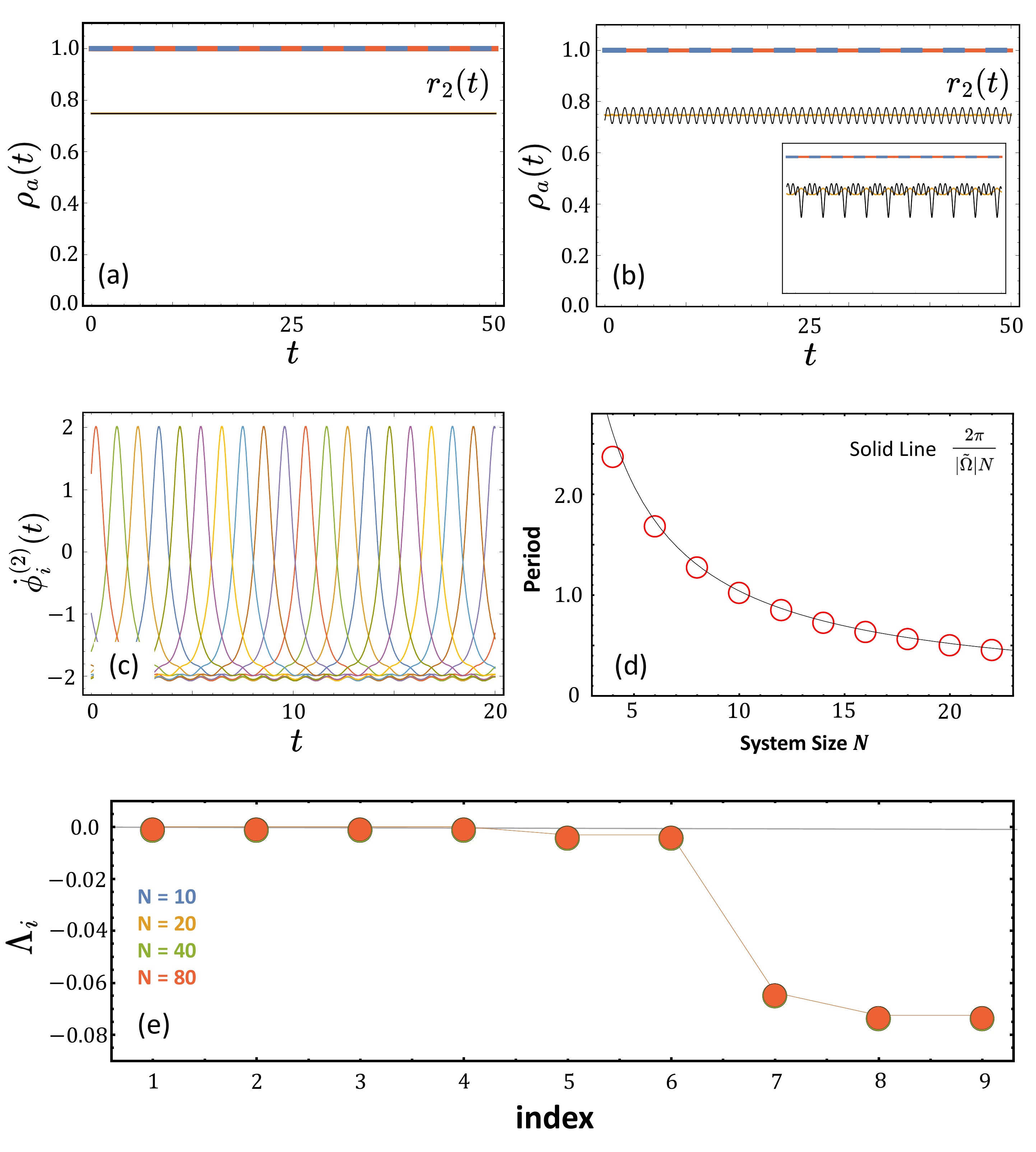}
\caption{Stationary SDS chimera states in the 9D Watanabe-Strogatz dynamics. (a,b) Time evolution of the radial macroscopic variables after a transient time of $10^5$ units for $A=0.35$. Blue and red lines indicate $\rho_1(t)=\rho_3(t)=1$ and the orange line $\rho_2(t)<1$. The black line shows the modulus of Kuramoto order parameter $r_2(t)$ calculated from the relation between Kuramoto order parameter and the WS variables with uniformly distributed constants of motion: (a) $N=10$ and (b) $N=40$. The inset of (b) shows the radial variables and Kuramoto order parameter with nonuniform constants of motion with $N=10$. (c) Instantaneous phase velocities obtained from Eq.~(\ref{eq:time-derivative}) for $N=10$ with uniform constants of motion. (d) Period of the modulus of the Kuramoto order parameter as a function of system size $N$. The red circles are numerically obtained periods and the black solid line is the curve $\frac{2\pi}{|\tilde{\Omega}|N}$. (e) Lyapunov exponents of the 9D macroscopic variables of the WS dynamics.} 
\label{Fig:SDS-WS}
\end{figure}

A bifurcation diagram of the stationary SDS chimera state is depicted in Fig.~\ref{Fig:SDS-OA} (c). It follows the same bifurcation scenario as chimeras either in two-population networks~\cite{abrams_chimera2008,abrams_chimera2016} or in three-population networks restricted to the symmetry-reduced manifold~\cite{martens_three}. They are created/destroyed in a limit point (LP) bifurcation, creating a stable and an unstable SDS branch. Following the stable branch to large values of $A$, it is destabilized in a supercritical Hopf bifurcation (HB) generating a stable breathing SDS chimera state. This breathing chimera disappears in a homoclinic bifurcation as the parameter $A$ is further increased (Compare this to Fig.~5 (a) in Ref.~\cite{martens_three}). 

In Fig.~\ref{Fig:SDS-WS}, we explore the 9D WS dynamics (\ref{eq:WS-governing-eq}) for the stationary SDS chimeras with uniform constants of motion. The macroscopic WS dynamics displays a prominent dependence on system size $N$. For large $N$, the radial variables (Fig.~\ref{Fig:SDS-WS} (a)) are stationary and characterized by $\rho_1(t)=\rho_3(t)=1$ (red and blue) and $\rho_2(t)=\rho_0 <1$ (orange coinciding with black, see below). For this solution, we find the angular variable $\partial_t \Phi_a (t) = \Omega$ for $a=1,2,3$, which in fact have the same characteristic as in the 6D OA dynamics above, i.e., $\rho_0=0.74769$ and $\Omega=-2.13448$ for $A=0.35$. The 
other angular variables read $\partial_t \Psi_a(t)=0$ for $a=1,3$, and $\partial_t \Psi_2(t) = \tilde{\Omega} = -0.60369$. For small $N$ (Fig.~\ref{Fig:SDS-WS} (b)), the radial variables are slightly fluctuating around the values of the OA dynamics (orange not coinciding with black), however, the fluctuations are so small that they can be neglected. The dependence on the system size is more evident in the Kuramoto order parameter calculated from Eq.~(\ref{eq:relation-kuramoto-WS}). For small $N$, the modulus of the Kuramoto order parameter of the incoherent population shows an oscillating motion along the $\rho_0$ value  (black, in Fig.~\ref{Fig:SDS-WS} (b)). Such a secondary oscillation disappears as the system size increases (black, in Fig.~\ref{Fig:SDS-WS} (a)), similar to Fig. 4 (a) in Ref.~\cite{lee1}: Both the amplitude and the period are decreasing as $N$ grows. This phenomenon can be understood as follows. First, we use the following form of the Watanabe-Strogatz transformation~\cite{WS_mobius,pikovsky_WS2,WS_original2}
\begin{flalign}
    \tan \Bigg( \frac{\phi^{(a)}_j-\Phi_a}{2} \Bigg) = \frac{1-\rho_a}{1+\rho_a} \tan \Bigg( \frac{\psi^{(a)}_j-\Psi_a}{2} \Bigg) \label{eq:WS transformation-original}
\end{flalign} for $a=1,2,3$ and $j=1,...,N$. This transformation directly gives a time derivative of each phase variable in terms of the WS global variables and the constants of motion. Starting with Eq.~(\ref{eq:WS transformation-original}) and its time derivative, we obtain
\begin{widetext}
\begin{flalign}
\frac{d}{dt}(\phi_j-\Phi) &= 2\frac{d}{dt} \Bigg[ \tan^{-1}\bigg( g \tan\bigg(\frac{\psi_j-\Psi}{2} \bigg)  \bigg)  \Bigg] =\frac{2}{1+g^2 \tan^2\big(\frac{\psi_j-\Psi}{2} \big)} \Bigg[ \dot{g}\tan\big(\frac{\psi_j-\Psi}{2}\big) + \frac{g}{\cos\big(\frac{\psi_j-\Psi}{2}\big)} \big(-\frac{1}{2}\dot{\Psi}\big) \Bigg] \notag \\
&=\frac{g}{\cos^2\big(\frac{\psi_j-\Psi}{2} \big)+ g^2 \sin^2\big(\frac{\psi_j-\Psi}{2} \big)} \Bigg( \frac{\dot{g}}{g}\sin(\psi_j-\Psi) - \dot{\Psi}  \Bigg) \label{sub1}
\end{flalign}
\end{widetext} for $j=1,...,N$ and $g:=\frac{1-\rho}{1+\rho} $. The denominator can be written, using the trigonometric identity, as
\begin{widetext}
\begin{flalign}
&\cos^2\big(\frac{\psi_j-\Psi}{2} \big)+ g^2 \sin^2\big(\frac{\psi_j-\Psi}{2} \big) = \frac{1+\cos(\psi_j-\Psi)}{2} + g^2 \frac{1-\cos(\psi_j-\Psi)}{2} \notag \\
&=\frac{1+\rho^2}{(1+\rho)^2} +\frac{2\rho}{(1+\rho)^2} \cos(\psi_j-\Psi) = \frac{1+\rho^2+2\rho\cos(\psi_j-\Psi)}{(1+\rho)^2} \label{sub2}
\end{flalign}
\end{widetext}The time derivative of $g$ is
\begin{equation}
\frac{\dot{g}}{g} = \frac{-\dot{\rho}(1+\rho)-\dot{\rho}(1-\rho)}{(1+\rho)^2}\frac{1+\rho}{1-\rho} = \frac{-2\dot{\rho}}{1-\rho^2}. \label{sub3}
\end{equation} Substituting Eqs.~(\ref{sub2}-\ref{sub3}) into Eq.~(\ref{sub1}), the phase velocity can be written as
\begin{flalign}
\dot{\phi}_j = \dot{\Phi} &+ \frac{1-\rho^2}{1+\rho^2+2\rho\cos(\psi_j-\Psi)} \notag \\
&\times \Bigg( \frac{-2\dot{\rho}}{1-\rho^2}\sin(\psi_j-\Psi) - \dot{\Psi} \Bigg) \label{sub4}
\end{flalign} for $j=1,...,N$. Hence, we obtain the instantaneous phase velocity of each oscillator 
\begin{widetext}
\begin{flalign}
    \dot{\phi}^{(a)}_j(t) = \dot{\Phi}_a - \frac{1-\rho_a^2}{1+\rho_a^2+2\rho_a\cos(\psi^{(a)}_j-\Psi_a)}  \Bigg( \dot{\Psi}_a 
    +\frac{2\dot{\rho}_a}{1-\rho_a^2}\sin(\psi^{(a)}_j-\Psi_a) \Bigg). \label{eq:time-derivative}
\end{flalign}
\end{widetext}Plugging the macroscopic variables $\rho_2 = \rho_0$, $\dot{\Phi}_2(t)=\Omega$, $\Psi_2(t) = \tilde{\Omega}t+\Psi_2(0)$ and the constants of motion $\psi^{(a)}_j$ uniformly distributed in $[-\pi,\pi]$ into Eq.~(\ref{eq:time-derivative}), we obtain
\begin{flalign}
     \dot{\phi}^{(2)}_j(t) &= \Omega - \frac{\tilde{\Omega}(1-\rho_0^2)}{1+\rho_0^2+2\rho_0\cos(\psi^{(2)}_j-\tilde{\Omega}t -\Psi_2(0))}   \notag \\
     &= \Omega - \frac{\tilde{\Omega}(1-\rho_0^2)}{1+\rho_0^2+2\rho_0\cos(\psi^{(2)}_j-\tilde{\Omega}(t-\frac{2\pi}{\tilde{\Omega}}) +\Psi_2(0))} \notag \\
     & = \dot{\phi}_j^{(2)}(t-T)  ~~ ~~ \text{where}~~T := \frac{2 \pi}{|\tilde{\Omega}|} \notag
\end{flalign} for $j=1,...,N$.
This indicates that the instantaneous phase velocity of each oscillator is a periodic function with the period $T = \frac{2 \pi}{|\tilde{\Omega}|}$. Furthermore, all the oscillators have the same functional form since they are determined by the same three WS variables ($\rho_0,\Omega$ and $\tilde{\Omega}$), and they are equally spaced within the time interval $T$ due to the uniform constants of motion (Fig.~\ref{Fig:SDS-WS} (c)). From this fact, we can assume $\dot{\phi}^{(2)}_{i}(t-\frac{j}{N}T)=\dot{\phi}^{(2)}_{i+j}(t)$ for an arbitrary $j \in \{1,...,N\}$, which gives $\phi^{(2)}_{i}(t-\frac{1}{N}T) = \phi^{(2)}_{i+1}(t)+C$ for $i=1,...,N$ with $\phi^{(2)}_{N+1} \equiv \phi^{(2)}_{1}$ where $C \in \mathbb{R}$ is a constant. This assumption~\cite{lee1,lee2} leads to
\begin{flalign}
    r_2(t) &= |\Gamma_2(t)| = \bigg| \frac{1}{N}\sum_{k=1}^{N}e^{i \phi_{k+1}^{(2)}(t)} \bigg| \notag \\
    &= \bigg| \frac{e^{-i C}}{N}\sum_{k=1}^{N}e^{i \phi_{k}^{(2)}(t-\frac{T}{N})} \bigg| =  \bigg| \frac{1}{N}\sum_{k=1}^{N}e^{i \phi_{k}^{(2)}(t-\frac{T}{N})} \bigg| \notag \\
    &=r_2\bigg(t-\frac{T}{N}\bigg) = r_2\bigg(t-\frac{2\pi}{| \tilde{\Omega} | N}\bigg) 
\end{flalign} where the period $ \frac{2\pi}{| \tilde{\Omega} | N}$ is decreasing as $N$ increases. In Fig.~\ref{Fig:SDS-WS} (d), we numerically measure the periods of the modulus of the Kuramoto order parameter for different sizes $N$ of the incoherent population (red circles) and compare them with $ \frac{2\pi}{| \tilde{\Omega} | N}$. The good agreement between the two curves evidences that the stationary SDS chimera state in a small-sized system continuously approaches the OA dynamics as $N \rightarrow \infty$ in a similar way found for the two-population networks~\cite{lee1}. The stationary SDS chimeras strongly depend on the constants of motion (equivalently, an initial condition of the microscopic dynamics).
The inset of Fig.~\ref{Fig:SDS-WS} (b), shows the temporal evolution of the radial WS variables and the Kuramoto order parameter for slightly nonuniform constants of motion. Clearly, the time series exhibit non-Poisson chimera features typical for chimeras outside the Poisson submanifold~\cite{pikovsky_WS1,lee1}.

To determine the stability of these SDS chimeras, we determined the Lyapunov exponents (Fig.~\ref{Fig:SDS-WS} (e)). Regardless of system size $N$, the stationary SDS chimeras in 9D WS dynamics are neutrally stable. Two of the zero Lyapunov exponents arise from the two continuous symmetries: the time shift and the phase shift invariance. There are two further zero exponents while the remaining Lyapunov exponents are negative. Thus, the SDS chimeras of the 9D WS dynamics are neutrally stable. For the breathing SDS chimeras, they follow the same dynamical and spectral properties as the breathing chimeras in two-population networks~\cite{abrams_chimera2008,abrams_chimera2016}, i.e. they possess an additional zero Lyapunov exponent due to the Hopf frequency.

Finally, we explore the $3N$-dimensional microscopic dynamics. To this end, two different initial conditions are exploited~\cite{lee1,martens_three}. A PIC (Poisson initial condition) is obtained by determining a fixed point solution of the 2D reduced OA equations in Eq.~(13) in Ref.~\cite{martens_three}, $\rho_0$ and $\varphi_0$. Then, the initial phases of the incoherent population are obtained by
\begin{equation}
    \frac{i-\frac{1}{2}}{N} = \frac{1}{2\pi} \int_{-\pi}^{\phi^{(a)}_i(0)}  \frac{1-\rho_0^2}{1-2\rho_0\cos(\phi-\varphi_0)+\rho_0^2} d\phi \notag
\end{equation} for $i=1,...,N$. The initial phases of the synchronized population are selected from a delta distribution. In contrast, an n-PIC (non-Poisson initial condition), i.e., a random initial condition, is generated by randomly selecting phases for all three populations from the uniform distribution of $[-\pi,\pi)$.

\begin{figure}[t!]
\includegraphics[width=1.0\linewidth]{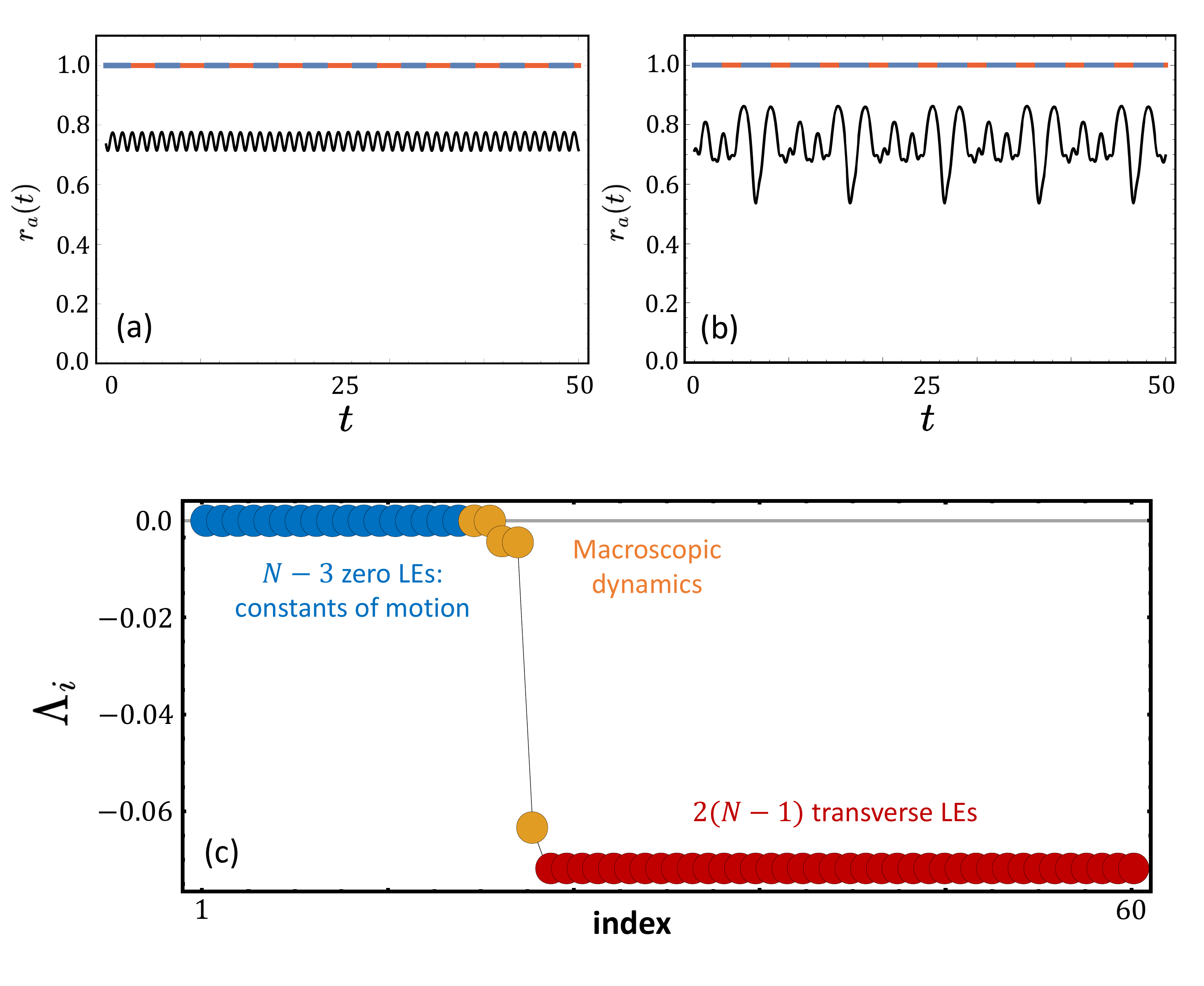}
\caption{Stationary SDS chimera states in $3N$-dimensional microscopic dynamics. (a,b) Time series of the moduli of the Kuramoto order parameters of the three populations with (a) PIC and (b) n-PIC for $N=10$ and $A=0.35$ after a transient time of $10^5$ units. (c) Lyapunov exponents of the SDS chimeras initiated from a PIC for $N=20$ and $A=0.35$. } 
\label{Fig:SDS-micro}
\end{figure}

In Fig.~\ref{Fig:SDS-micro} (a) and (b), the moduli of Kuramoto order parameters calculated from the microscopic dynamics are depicted for a PIC and an n-PIC, respectively, with $N=10$. For a PIC, the order parameter dynamics shows the same behavior as in Fig.~\ref{Fig:SDS-WS} (b), reflecting that a PIC corresponds to uniform constants of motion~\cite{abrams_chimera2016}. As the system size $N$ increases, the amplitude and the period of the secondary oscillation are decreasing and approach the OA dynamics for the same reason as discussed above for the WS dynamics. On the other hand, an n-PIC in Fig.~\ref{Fig:SDS-micro} (b) gives a non-Poisson chimera motion qualitatively similar to the inset of Fig.~\ref{Fig:SDS-WS} (a), corresponding to nonuniform constants of motion.

As far as Lyapunov stability is concerned, both PICs and n-PICs exhibit the same spectral characteristics in Fig.~\ref{Fig:SDS-micro} (c). Stationary SDS chimeras are neutrally stable with $N-1$ zero Lyapunov exponents, which explains their strong dependence on the initial condition (or constants of motion). $N-3$ of the zero Lyapunov exponents (blue) arise from the $N-3$ constants of motion, the remaining two (orange, zero) are expected to arise from the macroscopic dynamics. The two nearly identical and negative macroscopic Lyapunov exponents (orange) describe the stability with respect to perturbation along the two  synchronized populations, the strongly negative one (orange) is related to the WS radial variable of the incoherent population~\cite{lee1}. Moreover, there are $2(N-1)$-fold degenerate Lyapunov exponents (red). The covariant Lyapunov vectors (CLVs)~\cite{CLV1,CLV2,CLV3} corresponding to these Lyapunov exponents, given by
\begin{flalign}
    & \delta x_{\text{trans}} = (\delta a_1,...,\delta a_N,0,...,0,\delta b_1,...,\delta b_N)^\top \notag \\
    & \text{where}~~\sum_{k=1}^{N}\delta a_k = \sum_{k=1}^{N}\delta b_k =0, 
\end{flalign} reveal that these Lyapunov exponents determine the stability transverse to the synchronized populations.


\subsection{\label{subsec:DSD}Symmetric Chimera States of DSD-type}

\begin{figure}[t!]
\includegraphics[width=1.0\linewidth]{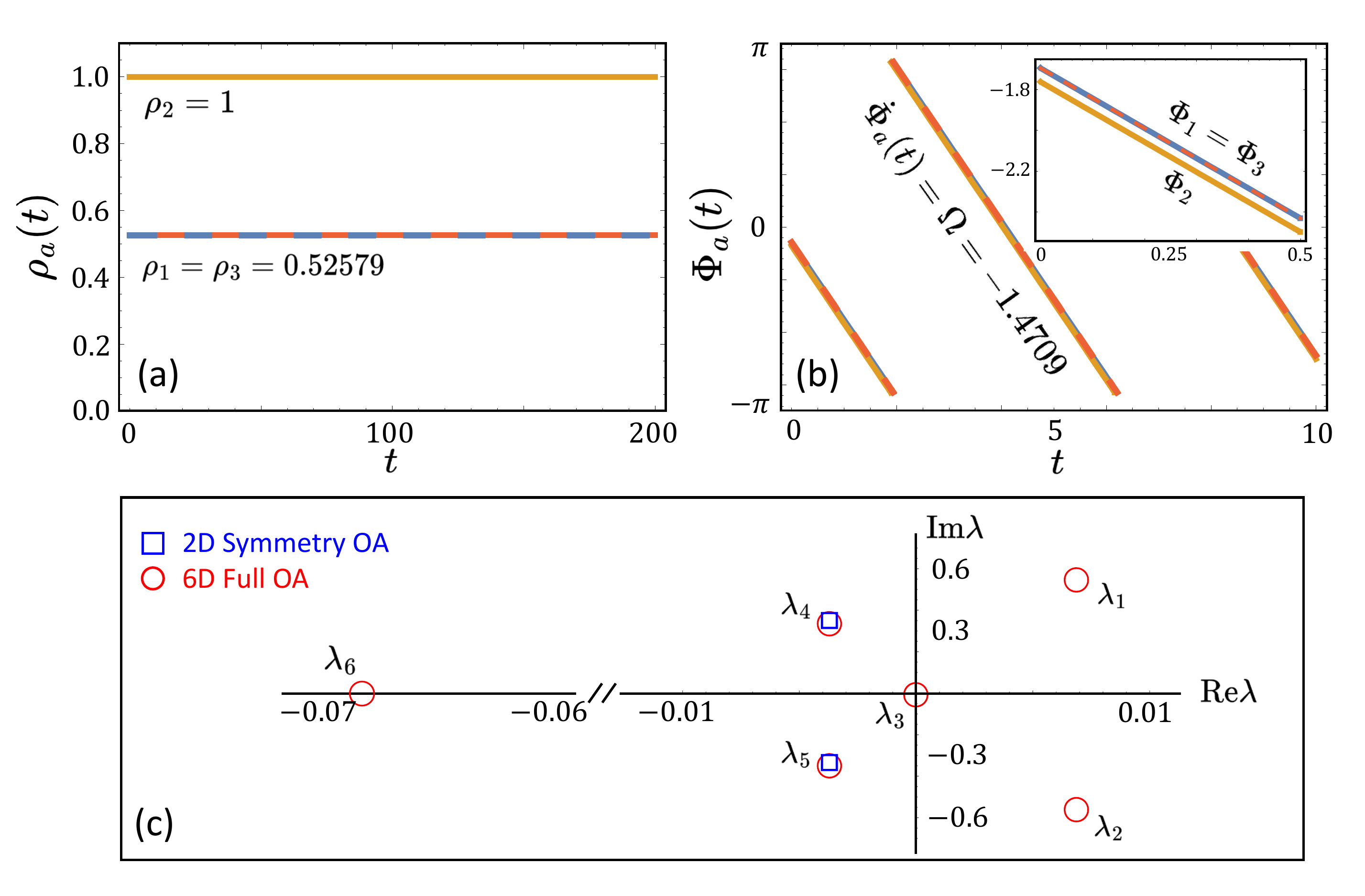}
\caption{ Unstable stationary DSD chimera state in 6D OA dynamics. (a) Radial variables $\rho_2(t)=1$ (orange line), $\rho_1(t)=\rho_3(t) = \rho_0 <1$ (red and blue lines) as a function of time for $A = 0.55$. (b) Time series of the angular variables with the same color scheme as used in (a). In (a) and (b) the first $10^5$ time units were discarded. (c) Eigenvalues of the Jacobian matrix evaluated at the DSD solution in the rotating reference frame.  } 
\label{Fig:DSD-OA}
\end{figure}

Figure~\ref{Fig:ensemble-table} shows that in none of the simulations a stationary or breathing DSD symmetric chimera state was observed. These states will turn out to be unstable later in this section. To investigate their dynamical and spectral properties, we need a very specific initial condition. In the 6D OA dynamics, we again look for a symmetric fixed point solution of the OA equations obeying $\rho_1(0)=\rho_3(0)=\rho_0 <1$, $\rho_2(0)=1$, $\Phi_1(0)=\Phi_3(0) = \varphi_0 \in \mathbb{T}$ and $\Phi_2(0)=0$. Here, $\rho_0$ and $\varphi_0$ are the stable symmetric DSD solutions in the reduced manifold (see Eq.~(14) in Ref.~\cite{martens_three}). In Fig.~\ref{Fig:DSD-OA} (a-b), a stationary DSD symmetric chimera state from such an initial condition is shown. It is characterized by $\rho_1(t)=\rho_3(t) = \rho_0 <1$, $\rho_2(t)=1$, $\Phi_1(t)=\Phi_3(t)=\Omega t + \varphi_0$ and $\Phi_2(t)=\Omega t$. For $A=0.55$ where a stable stationary DSD symmetric chimera can be found in the reduced manifold, the numerical values are $\rho_0=0.52579$ and $\Omega = -1.4709$. Likewise, in a rotating reference frame with $\Omega$, it is a fixed point solution, which allows us to perform a linear stability analysis and determine the eigenvalues of the Jacobian matrix.

In Fig.~\ref{Fig:DSD-OA} (c), the eigenvalues are depicted in the complex plane. Two of them have positive real parts: $\lambda_1 = \overline{\lambda}_2$ with corresponding eigenvectors $\delta x_1 = \overline{\delta x}_2 = (\delta a, 0, -\delta a, \delta b, 0, -\delta b)^\top$ for $\delta a \in \mathbb{C}$ and $\delta b \in \mathbb{R}$. This demonstrates that the unstable directions are transverse to the symmetry-reduced manifold, i.e., the first and the third populations are opposite to each other. The eigenvector corresponding to $\lambda_3=0$ is given by $\delta x_3 = (0,0,0,\delta a,\delta a,\delta a)^\top$ with $\delta a = 1/\sqrt{3}$, and thus describes a common phase shift. The perturbations corresponding to the pair of complex conjugate eigenvalues $\lambda_4 = \overline{\lambda}_5$ are parallel to the symmetry-reduced manifold: $\delta x_4 = \overline{\delta x}_5 = (\delta a,0, \delta a, \delta b, \delta c, \delta b)^\top$ for $\delta a, \delta b \in \mathbb{C}$ and $\delta c \in \mathbb{R}$. Finally, $\delta x_6 = (\delta a ,\delta b, \delta a, \delta c,\delta c,\delta c)^\top$ for $\delta a, \delta b, \delta c \in \mathbb{R}$ points in the radial direction parallel to the reduced manifold with a common phase shift for all three populations.

On the level of the WS dynamics with uniform constants of motion, we also require a specific initial condition to study the unstable stationary DSD symmetric chimera state that can be obtained in the symmetry-reduced WS dynamics: $\rho_1=\rho_3 =: \rho<1$, $\rho_2=1$, $\varphi := \Phi_1-\Phi_2=\Phi_3-\Phi_2$ and $\Psi_1=\Psi_3$ with $\Psi := \Psi_1 - \Psi_2=\Psi_3 - \Psi_2$ governed by
\begin{flalign}
    \dot{\rho} = \frac{1-\rho^2}{2}\big( &\mu(\zeta \cos \alpha  +\xi \sin \alpha) -\nu\sin(\varphi +\alpha) \notag \\
    & +\nu (-\zeta \sin\alpha + \xi \cos\alpha) \big) \notag \\
    \dot{\Psi} = \frac{1-\rho^2}{2\rho} \big(  &\mu (-\zeta \sin\alpha+\xi\cos\alpha) -\nu\sin(\varphi+\alpha) \notag \\
    &+\nu(-\zeta\sin\alpha+\xi\cos(\varphi-\alpha))\big) \notag \\
    \dot{\varphi} = \frac{1+\rho^2}{2\rho} \big( &\mu(-\zeta\sin\alpha+\xi\cos\alpha)-\nu\sin(\varphi+\alpha) \notag \\
    &\nu(-\zeta\sin\alpha+\xi\cos\alpha)  \big) -\big( -\mu\sin\alpha \notag \\
    &+2\nu(\zeta\sin(\varphi-\alpha)+\xi\cos(\varphi-\alpha)) \big)  \notag
\end{flalign}
where
\begin{flalign}
    \zeta &= \frac{1}{N}\sum_{k=1}^{N}\frac{2\rho + (1+\rho^2)\cos(\psi_k-\Psi)}{1+2\rho\cos(\psi_k-\Psi)+\rho^2}, \notag \\
    \xi &= \frac{1}{N}\sum_{k=1}^{N}\frac{ (1-\rho^2)\sin(\psi_k-\Psi)}{1+2\rho\cos(\psi_k-\Psi)+\rho^2} \notag
\end{flalign} for $\psi_k = -\pi +(2\pi) \frac{k-1}{N}$, $k=1,...,N$. In this reduced system, the symmetric DSD chimeras are found to be stable. Then, we use $\rho_1(0)=\rho_3(0)=\rho(T), \rho_2(0)=1$, $\Phi_1(0)=\Phi_3(0)=0, \Phi_2(0)=\varphi(T)$, and $\Psi_1(0)=\Psi_3(0)=\Psi(T), \Psi_2(0) \in \mathbb{T}$ as an initial condition of Eq.~(\ref{eq:WS-governing-eq}) for $T \gg 1$.

\begin{figure}[t!]
\includegraphics[width=1.0\linewidth]{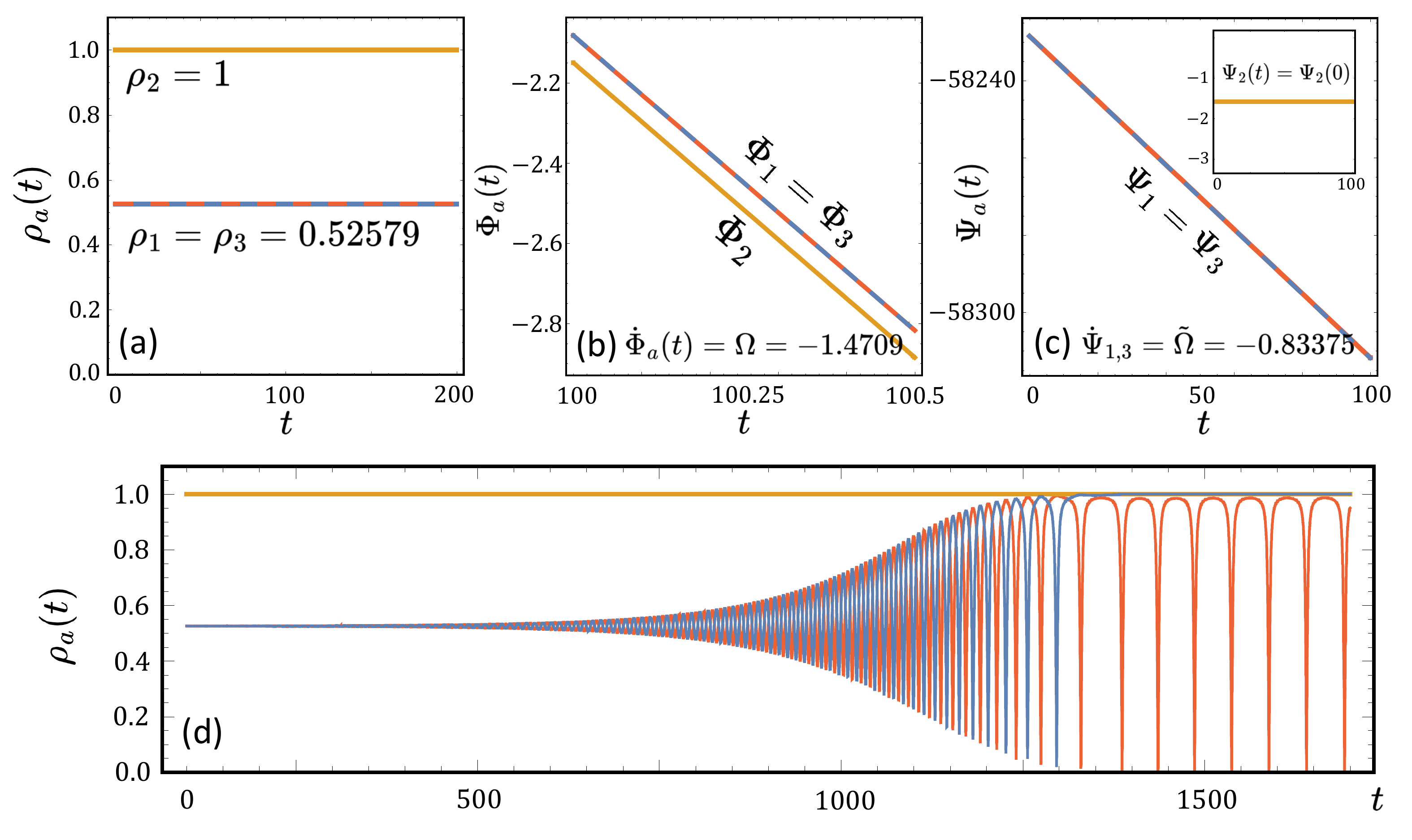}
\caption{ Unstable stationary DSD chimera states in 9D WS dynamics. (a-c) Time evolution of the 9D Watanabe-Strogatz macroscopic variables after a transient time of $10^5$ units with the same color scheme in Fig.~\ref{Fig:DSD-OA}. (d) Time evolution of the radial variables with a small perturbation on the specific initial condition. $A=0.55$ and $N=40$. } 
\label{Fig:DSD-WS}
\end{figure}

\begin{figure}[t!]
\includegraphics[width=1.0\linewidth]{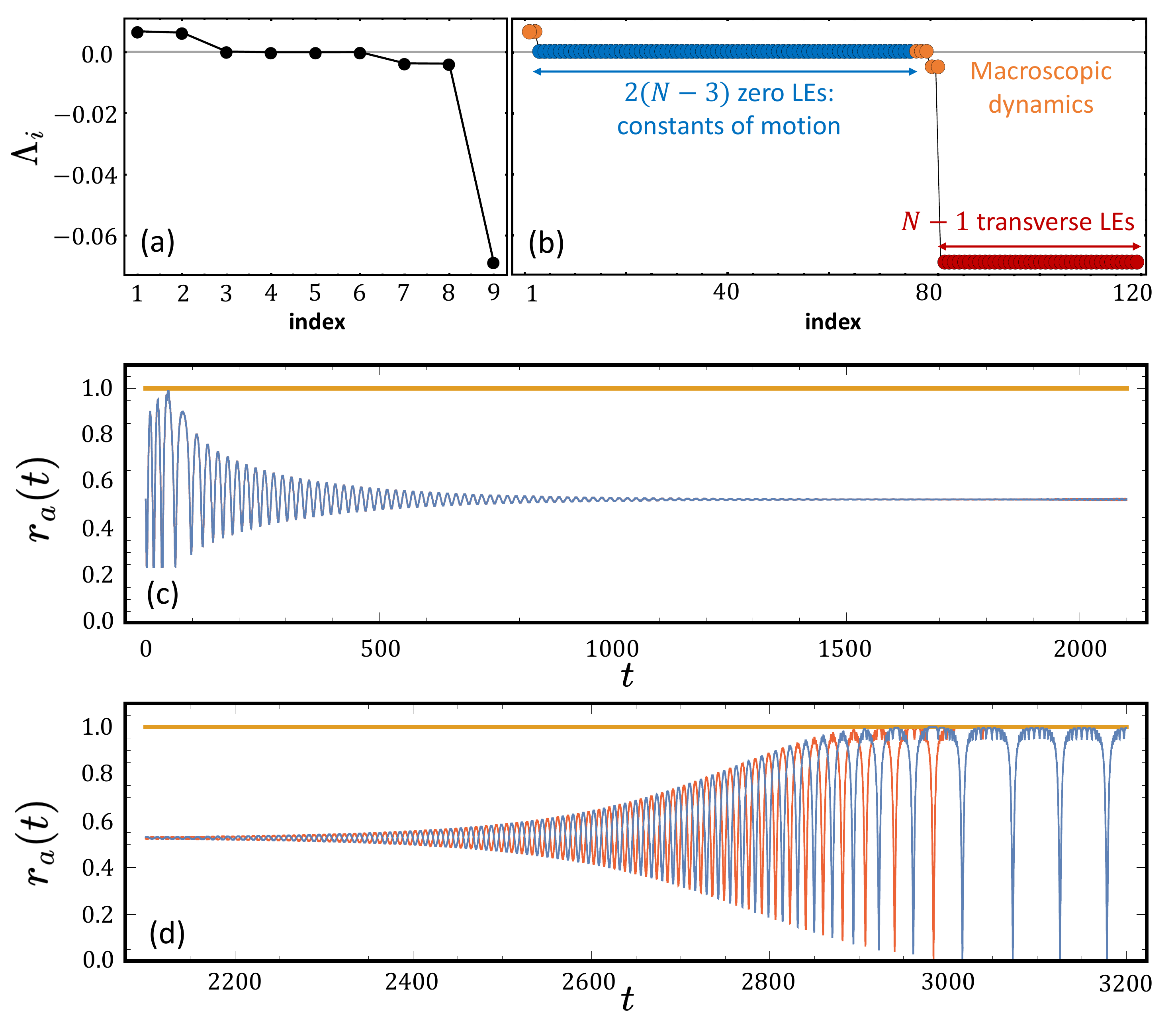}
\caption{(a) Lyapunov exponents of the 9D WS dynamics for $N=40$ corresponding to Fig.~\ref{Fig:DSD-WS} (a-c). (b) Lyapunov exponents of the $3N$-dimensional microscopic dynamics for $N=40$. (c-d) Time evolution of the moduli of the Kuramoto order parameters for the $3N$-dimensional microscopic dynamics with $N=40$ starting from a PIC. $A=0.55$.} 
\label{Fig:DSD-micro}
\end{figure}

Figure~\ref{Fig:DSD-WS} (a-c) displays the temporal dynamics of such a DSD symmetric chimera when starting (within our numerical resolution) directly on the unstable state. However, imposing a minor perturbation on these initial conditions, the stationary symmetric DSD chimera state occurs only as a transient, and evolves then through an antiphase oscillation to a breathing SDS chimera state (Fig.~\ref{Fig:DSD-WS} (d)). The Lyapunov exponents of the stationary DSD chimeras in the 9D WS dynamics are shown in Fig.~\ref{Fig:DSD-micro} (a). There are two positive Lyapunov exponents: $\Lambda_1$ and $\Lambda_2$. This does not indicate a chaotic attractor since any symmetric DSD chimera does not involve any chaotic motion and thus should be considered as an unstable solution. Moreover, the CLVs of them have the form $\delta x_{1,2} = (\delta a,0,-\delta a, \delta b,0,-\delta b, \delta c,0,-\delta c)$. This also confirms that the unstable directions of the stationary symmetric DSD chimera state are transverse to the symmetry-reduced manifold, i.e., the perturbations of population one and three are opposite to each other. This fact explains why one can obtain symmetric DSD chimeras in the reduced manifold but not in the full dynamics.

For the microscopic dynamics, even when starting from a precise PIC (equivalently, uniform constants of motion), we obtain a strange transient. First the system apparently approaches the stationary DSD state in an oscillatory manner and resides close to it for some hundred time units, before it attains a transient antiphase motion and finally reaches a breathing SDS chimera state (Fig.~\ref{Fig:DSD-micro} (c-d)).
Since in the microscopic dynamics, we were not able to observe a symmetric DSD chimera that lives long enough to investigate the Lyapunov stability, we detour to the 9D WS variables $\rho_a(t)$, $\Psi_a(t)$ and $\Phi_a(t)$ and uniform constants of motion together with the inverse WS transformation from Eq.~(\ref{eq:WS transformation-original}):
\begin{equation}
    \phi_j^{(a)}(t) = \Phi_a(t) +2 \tan^{-1} \Bigg( \frac{1-\rho_a(t)}{1+\rho_a(t)}\tan \bigg(\frac{\psi_j^{(a)}-\Psi_a(t)}{2}\bigg) \Bigg) 
    \notag
\end{equation} for $j=1,...,N$ and $a=1,2,3$. First, we implement a numerical integration of the 9D WS equations with the very initial condition for the stationary DSD chimeras. Then, we obtain the macroscopic variables as a function of time, which are plugged into the above transformation. This gives the time evolution of each phase variable. Then, the tangent space dynamics is governed by the Jacobian matrix defined by
\begin{flalign}
     (\mathbf{J})_{ij}(t) = \begin{pmatrix}
    \frac{\partial \dot{\phi}^{(1)}_i }{\partial \phi^{(1)}_j} &\vline & \frac{\partial \dot{\phi}^{(1)}_i}{\partial \phi^{(2)}_j} & \vline & \frac{\partial \dot{\phi}^{(1)}_i}{\partial \phi^{(3)}_j}   \\ \cline{1-5}
    \frac{\partial \dot{\phi}^{(2)}_i }{\partial \phi^{(1)}_j} &\vline & \frac{\partial \dot{\phi}^{(2)}_i}{\partial \phi^{(2)}_j} & \vline & \frac{\partial \dot{\phi}^{(2)}_i}{\partial \phi^{(3)}_j} \\ \cline{1-5}
    \frac{\partial \dot{\phi}^{(3)}_i }{\partial \phi^{(1)}_j} &\vline & \frac{\partial \dot{\phi}^{(3)}_i}{\partial \phi^{(2)}_j} & \vline & \frac{\partial \dot{\phi}^{(3)}_i}{\partial \phi^{(3)}_j}
    \end{pmatrix}
\end{flalign} evaluated at $\phi_j^{(a)}(t)$ above for $i,j=1,...,3N$. The tangent linear propagator is defined by $\mathbf{M}(t,t_0) = \mathbf{O}(t)\mathbf{O}(t_0)^{-1}$ where $\dot{\mathbf{O}}(t) = \mathbf{J}(t)\mathbf{O}(t)$ with $\mathbf{O}(0)=I_{3N}$~\cite{CLV1,CLV2,CLV3}. Then, we obtain the Lyapunov exponents as 
\begin{equation}
    \Lambda_i = \lim_{t\rightarrow \infty} \frac{1}{t}\log \frac{||\mathbf{M}(t,t_0)\mathbf{\delta u}_i(t_0)||}{||\mathbf{\delta u}_i(t_0)||} \notag
\end{equation} where $\mathbf{\delta u}_i(t_0)$ belongs to each Oseledets' splitting for $i=1,...,3N$~\cite{oseledets,pikovsky_LE}.

In Fig.~\ref{Fig:DSD-micro} (b), we show the Lyapunov exponents of $3N$-dimensional microscopic dynamics as obtained with the above numerical scheme. There are two positive exponents with CLVs of the form
\begin{flalign}
    & \delta x_{1,2} =(\delta a_1,...,\delta a_N,0,...,0,\delta b_1, ...,\delta b_N)^\top  \notag \\
    &\text{where} ~~ \delta b_i = -\delta a_i \notag
\end{flalign} for $i=1,...,N$, which again elucidates that the unstable directions are transverse to the symmetry-reduced manifold. Also, there are $2(N-3)$ zero Lyapunov exponents (blue) from the constants of motion for the two incoherent populations. The CLVs corresponding to the $(N-1)$-fold degenerate transverse Lyapunov exponents (red) are given by
\begin{flalign}
    & \delta x_{\text{trans}} = (0,...,0,\delta a_1,...,\delta a_N,0,...,0)^\top \notag \\
    &\text{where} ~~\sum_{k=1}^N \delta a_k = 0. \label{eq:clvs-dsd-transverse}
\end{flalign} The other Lyapunov exponents including three zero-valued ones (orange) are expected to occur from the WS macroscopic variables and perturbations along the synchronized population.

\bibliography{apssamp}

\end{document}